\documentclass{article}

\usepackage{PRIMEarxiv}

\usepackage[utf8]{inputenc} % allow utf-8 input
\usepackage[T1]{fontenc}    % use 8-bit T1 fonts
\usepackage{hyperref}       % hyperlinks
\usepackage{url}            % simple URL typesetting
\usepackage{booktabs}       % professional-quality tables
\usepackage{amsfonts}       % blackboard math symbols
\usepackage{nicefrac}       % compact symbols for 1/2, etc.
\usepackage{microtype}      % microtypography
\usepackage{lipsum}
\usepackage{fancyhdr}       % header
\usepackage{graphicx}       % graphics
\graphicspath{{media/}}     % organize your images and other figures under media/ folder

\title{\texttt{ProvDeploy}: Provenance-oriented Containerization of High Performance Computing Scientific Workflows
\thanks{\textit{\underline{Citation}}: 
\textbf{ Liliane Kunstmann, Débora Pina, Daniel de Oliveira, Marta Mattoso. \texttt{ProvDeploy}: Provenance-oriented Containerization of High Performance Computing Scientific Workflows}} 
}

\author{
  Liliane Kunstmann, Débora Pina \\
  Department of Computer Science, COPPE \\
  Federal University of Rio de Janeiro \\
Rio de Janeiro, Brazil\\
  \texttt{\{lneves, dbpina\}@cos.ufrj.br} \\
   \And
  Daniel de Oliveira \\
  Instituto de Computação \\
  Fluminense Federal University \\
  Rio de Janeiro, Brazil\\
  \texttt{danielcmo@ic.uff.br} \\
    \And
  Marta Mattoso \\
   Department of Computer Science, COPPE \\
  Federal University of Rio de Janeiro \\
  Rio de Janeiro, Brazil\\
  \texttt{marta@cos.ufrj.br} \\
}

\begin{document}
\maketitle

\begin{abstract}
Many existing scientific workflows require High Performance Computing environments to produce results in a timely manner. These workflows have several software library components and use different environments, making the deployment and execution of the software stack not trivial. This complexity increases if the user needs to add provenance data capture services to the workflow. This manuscript introduces \texttt{ProvDeploy} to assist the user in configuring containers for scientific workflows with integrated provenance data capture. \texttt{ProvDeploy} was evaluated with a Scientific Machine Learning workflow, exploring containerization strategies focused on provenance in two distinct HPC environments
\end{abstract}

% keywords can be removed

\section{Introduction}
Several scientific and commercial workflows demand High Performance Computing (HPC) \cite{DBLP:journals/access/EliaFA21} environments. Computational solutions have been proposed to support the development of workflows capable of processing and extracting useful information from these data. It is common for large-scale scientific workflows to make use of libraries for parallel processing (\textit{e.g.}, Dask\footnote{\url{https://www.dask.org/}}), code reuse, or even process platforms from streams (\textit{e.g.}, Kafka). This myriad of solutions ended up creating a complex software ecosystem, and HPC installations can no longer keep up with this accelerated growth of software programs and libraries\cite{yuan2020bioinformatics}. As a result, developers have to do extra (and often complex) work to deploy their workflows, \textit{e.g.}, install multiple software from source, change environment variables, install dependencies, and change system files, which can be an error-prone task (due to version incompatibilities, for example). Distinctively from standalone applications, workflows comprise multiple activities. Each workflow activity can be represented by a program that demands its software requirements. All the activities have to be in tune with the workflow to be deployed, and this is a challenge mainly in moving workflows between different HPC environments.

The scientific workflow execution process also requires using additional software to support debugging, profiling, and provenance data capture \cite{freire2008provenance} to support analysis, monitoring, and repetition of executions. Provenance data might help runtime analysis of intermediate data \cite{DBLP:conf/eScience/OcanaSOM15}, fault tolerance \cite{DBLP:journals/cluster/GuedesJODO20}, and especially the reproduction of HPC executions \cite{DBLP:journals/tpds/HarrellMM22}. However, scientists already have to manage a complex software stack with their deployment issues, and adding provenance means adding new dependencies to its software stack and requirements.

An alternative to addressing workflow deployment challenges in HPC environments, especially for scientific workflows, is the use of containers\cite{DBLP:conf/fogiot/StruharBAP20}. Containerization can be described as OS-level virtualization, with kernel sharing. Containers are stateless software applications, meaning that upon execution completion, any data generated inside the container will be lost. To prevent this issue, the user has to set up mechanisms for data persistence, such as attached volumes and directories. Despite its popularity and wide adoption, container usage in HPC is not simple\cite{DBLP:journals/cluster/LiuG22}. Container kernel sharing and privileges\cite{priedhorsky2021minimizing} are a source of concern in HPC facilities because a compromised container can expose the whole environment to cyberattacks. As a result, container adoption is usually restricted to some technologies, like Singularity and Shifter, functionalities, and repositories. Those restrictions added to kernel architecture differences between environments cause container images to be rebuilt for each HPC environment, and creating a similar container image is not always possible in HPC \cite{canon2020role,wofford2022complete}.

Moreover, it is still up to the user/developer to configure how the containers will be created (\textit{i.e.}, which programs, libraries, and frameworks will be in each container), which can be complex depending on the target workflow. There is a large number of options for the same workflow to be containerized, which we call a \textit{Containerization Strategy}. A containerization strategy involves defining the number of images, how workflow activities and the provenance service will be organized in them, and how the containers interact. This containerization strategy becomes more critical when provenance data are used to monitor and tune the execution of the workflow. Provenance data capture usually is a data-intensive service, that can affect the performance of workflow activities executing via containers. Therefore, depending on how the user configures the containers, various problems may arise, such as increasing communication between containers, image transfer, data sharing and build time. More specifically, software conflicts between workflow activities and the provenance service may have to be addressed considering the HPC environment. Furthermore, some workflow activities may be already containerized and the configuration might change accordingly.

The most common and simple strategy is to reproduce the whole workflow software stack in a single container image \cite{schulz2016use, elisseev2022multiscale}, named coarse-grained strategy, but, the attempt to replace workflow activities to explore different software becomes a difficult task. The opposite of this strategy would be defining a container image for each workflow activity, named fine-grained strategy. In this case, the overhead generated by using more container instances might compromise workflow performance. This strategy also requires external tools to control container execution. Between coarse and fine-grained strategies there are other combinations of container images, which we call hybrid strategies, that can also be useful in specific cases, but remain unexplored in the literature. Although both containers and provenance are important in workflow deployment and development, we have not found in the literature solutions that support multiple strategies to deploy scientific workflows with provenance data capture in HPC environments.

In this manuscript, we propose a lightweight framework named \texttt{ProvDeploy} to support the deployment of scientific workflows with provenance data capture in HPC environments allowing for multiple containerization strategies. \texttt{ProvDeploy} considers the entire software stack needed by both the scientific workflow and the provenance service at deployment with containers. We present \texttt{ProvDeploy} being used with different strategies with DenseED, a scientific machine learning workflow. The results of this evaluation point out that the number of container instances is not a determining factor for the overhead and hybrid strategies can be more beneficial for workflow isolation, reuse, and data sharing. The remainder of this manuscript is structured as follows.
Section \ref{refteo} presents background on containers, provenance, and related work. Section \ref{provdeploysec} presents the framework \texttt{ProvDeploy}. Section \ref{avexp} shows the use of \texttt{ProvDeploy} supporting different containerization strategies with DenseED in two HPC environments, and Section \ref{sec:conclusao} concludes this manuscript. 

\section{Background and related work}\label{refteo}
Containers are an important support for workflow deployment and repeatability. Capturing provenance data from containers is essential for reproducibility\cite{ahmad2022reproducible,kleinsteuber2024managing} and security\cite{abbas2022paced,chen2021clarion} in applications in general. However, combining workflow containerization and provenance capture is still an open issue, particularly in HPC environments. In this section, we provide background information on containers and provenance and discuss how related work addresses these two properties while deploying workflows in HPC.

\subsection{Containerization Principles}

The term ``container'' denotes a space for storing multiple things, like a shipping container. However, a software container is more than a storage repository, since this technology packs and unpacks software stacks to be executed in multiple computational environments. Containerization allows moving the execution of software between environments simply, as long as the environments provide the necessary support for execution through containers. Most of the containerization technologies are compatible (through \textit{runc}), but they usually are distinguished by their engines, such as Docker, Singularity, and CRI-O. The container engine is the component responsible for creating container images from a description provided by the user (\textit{e.g.} definition file, \textit{recipe}, or \textit{dockerfile}). A container image is an executable software package that contains source code, libraries, files, data, \textit{etc}. The description provided by the user to create an image should contain the software stack to be packed, its dependencies, and the files that are necessary for its execution. With this description, the container engine packs those elements using virtualization techniques to set up an image of the software stack, so it can be unpacked and executed at its destination. This image can be used to create new images or to execute isolated (containerized) processes.

Container images can be shared between users and on public registries such as Docker Hub. Much of the value in the Docker ecosystem and containerization comes from the ability to push and pull repositories and images from registry servers, allowing sharing and reuse of container images. However, container kernel sharing can introduce security risks to the environment they are executed. Additionally, container images, being immutable, can become vulnerable to cyberattacks over time. Docker, the most popular container engine requires privileged use and is not capable of exploring optimized software that may be in the HPC machine. As a result, it has not seen widespread adoption in HPC installations, leading to the emergence of HPC container solutions like Shifter \cite{gerhardt2017shifter}, CharlieCloud \cite{priedhorsky2017charliecloud}, and Singularity\cite{kurtzer2017singularity}. These solutions meet requirements for more customizable environments providing more autonomy to users on HPC facilities. They focus on the mobility of execution environments and mitigate the problems that kernel sharing might cause. They also allow using optimized hardware and software that are in HPC facilities. However, there are still problems related to the use of containers in HPC, such as permissions/functionalities in different HPC facilities, familiarity with the technology and deciding on the best ways of deploying it, compatibility of kernel architectures, and configuration of containers to run a specific workflow.

On the task of deploying commercial applications, containers have shown to be efficient\cite{cito2017empirical}. As stateless, standalone applications, containers can deploy commercial applications in small pieces that execute tasks independently. Due to its popularity, its use and best practices are already standardized for the cloud\footnote{\url{https://cloud.google.com/architecture/best-practices-for-building-containers}}. HPC workflows, though, have distinct requirements for containers and resources. They usually require managing data generation and sharing, the interconnection between components, and having an explicit execution order. Thus, managing workflow requirements with the characteristics of containers poses a significant challenge in deploying HPC workflows. 

The generation of data in containers requires setting up volumes or bind paths, according to the container technology, but sharing these data is still a challenge. Besides that, organizing workflows in containers also creates the need to manage communication between containers, which is required by the workflow. Also, container images that run scientific workflows and capture provenance will require awareness of the container and thus data persistence. In addition, container usage in HPC is limited by the execution environment which may allow the use of specific container engines, functionalities, and repositories. So the user has to adapt to what is provided by the facility. For example, image building is a functionality that is usually not allowed in HPC facilities so the user needs to build images in another environment.
\subsection{Provenance Services }

Provenance data can be defined as the description of the origin of a data element, the activity that generated it, and the agent of this activity \cite{moreau2013provenance}. So, provenance data refers to the derivation history of a set of results, including activities (or processes) that produced those results. An increase in data quality, support for data interpretation and analysis, and data reproduction are known as benefits of provenance data capture. Capturing provenance data requires adopting an external tool or service that may persist provenance data in logs, files, or DBMSs.

In addition to data capture, querying provenance data can be necessary at runtime \cite{MATTOSO2015100}. The execution of an HPC workflow may last for hours or days and this kind of query is usually required because it allows for execution monitoring through a data derivation path. Requirements for provenance data capture in HPC workflows are the possibility to capture data from multiple workflow activities, in a non-intrusive way with negligible overhead. Some of those workflows have specific functions and variables that the user will monitor if only this data is captured in a way that helps its analysis, the user will not have to spend time on post-processing. DfAnalyzer \cite{DBLP:journals/softx/SilvaCGCOCVM20} is an example of a tool that captures and manages provenance data, allows query processing at runtime, and provides asynchronous data persistence which does not affect the execution time of the workflow. Furthermore, DfAnalyzer allows the user to specify the data to be captured.

Provenance services, like DfAnalyzer, add complexity to workflow deployment by running parallel to the whole workflow, sending and receiving multiple calls from multiple workflow activities, and persisting data through logs or DBMSs.  Those challenges are similar to the use of containers in workflows. However, provenance services are connected to the execution of all activities but launched independently, they are not managed by the workflow and persist data continuously.

\subsection{Related Work}

In this section, we initially present the main limitations of the most popular container orchestrators to deploy workflows in HPC. Then, we discuss current approaches for deploying workflows in notebooks to be later executed in HPC. We then present container support for a few HPC workflow systems, and finally, we review provenance service providers for containers. However, we did not find any solution that, like \texttt{ProvDeploy}, supports different containerization strategies for workflows and captures provenance from containerized workflows in HPC.

Despite the popularity of Docker and Kubernetes, these container solutions present limitations in HPC environments\cite{balis2022auto,zhou2021container}. One of the issues is related to performance impact due to restarting the containers for workflow parallel execution. The work of Straesser et al.\cite{straesserempirical} undertakes an empirical study with Docker images to identify the image features (e.g., image size, number of volumes) that have the most impact on the container start time. Their findings indicate that there is no single feature that affects alone container start time; instead, these features have to be considered alongside hardware and software configuration.

Some approaches, though, combine Docker with provenance data services to support the execution and reproduction of scientific software applications \cite{DBLP:conf/sigmod/ChirigatiRSF16,that2017sciunits}. ReproZip\cite{DBLP:conf/sigmod/ChirigatiRSF16} is a pioneering tool that provides a reproducible artifact in different formats (e.g., zip, Docker, Vagrant). ReproZip allows users to automatically and transparently capture all the dependencies of a computational experiment in a single, distributable bundle (i.e., Research Object\cite{bechhofer2010research}), that can be used to reproduce the entire experiment in another environment, excluding HPC. ReproServer \cite{rampin2018reproserver}, an extension of ReproZip, is not intrusive or time-consuming to encourage verifying results. ReproSever is an open-source Web application that allows users to reproduce experiments tracked with ReproZip, from their Web browser, without downloading or installing software or data. However, they adopt coarse-grained containerization and are limited to Docker. 

Kubernetes is an open-source system for automating containerized application deployment, scaling, and management, particularly excelling in managing microservice applications. However, it is not designed for workflows or widely adopted in HPC facilities. To address Kubernetes’ limitations for HPC, solutions such as Balis et al.’s approach\cite{balis2022auto} leverage autoscaling, taking advantage of the known workflow structure to improve scaling decisions by predicting resource demands for the execution of each activity of the workflow. Conversely, the work of Shan et al. \cite{shan2023kubeadaptor} explores Kubernetes inconsistencies in the task scheduling order, which damages HPC workflow execution. Their proposed work, KubeAdaptor, is a cloud-based framework able to implement workflow containerization on Kubernetes and integrate workflow systems with Kubernetes, focusing on the consistency of task scheduling order. 

Zhou et al. \cite{zhou2021container} propose a TORQUE-operator that bridges Kubernetes and TORQUE scheduler providing support for microservices that HPC schedulers lack. This solution is also focused on hybrid architectures, providing a unique access interface for cloud and HPC jobs using Singularity on its Kubernetes cluster. Currently, Singularity\cite{kurtzer2017singularity} is the standard container engine for HPC facilities\cite{zhou2021container}, because it runs containers capable of exploring optimized libraries, has native support to MPI and GPUs, requires no daemon process, and runs containers with current user privileges. Singularity does not rely on Docker to create images but it can convert and execute Docker images. Kubernetes can support Singularity automation through Singularity-CRI since Singularity is OCI\footnote{\url{https://opencontainers.org/}}-compliant (Open Container Initiative -  has the purpose of creating open industry standards around container formats and runtimes). Many approaches like kube-batch\footnote{\url{https://github.com/kubernetes-retired/kube-batch}} and Hkube\footnote{\url{https://hkube.io/}} have been released to provide Kubernetes batch execution and scheduling. However, Kubernetes has now native features for that purpose, Jobs\footnote{\url{https://kubernetes.io/docs/concepts/workloads/controllers/job/}} for batch execution and CronJob\footnote{\url{https://kubernetes.io/docs/concepts/workloads/controllers/cron-jobs/}} for job scheduling. A Kubernetes Job will execute one or more containers until execution successfully terminates for a specific number of containers and CronJob can schedule multiple Jobs. There are multiple efforts to integrate and increase the adoption of Kubernetes in HPC workloads and environments, because of its efficiency for container automation.

In addition to the approaches mentioned, there are Kubernetes-based solutions that aim to support workflow execution such as Pachyderm\cite{novella2019container}\footnote{\url{https://www.pachyderm.com/}}, Kubeflow\cite{bisong2019kubeflow} \footnote{\url{https://www.kubeflow.org/}}, Polyaxon\footnote{\url{https://polyaxon.com/}}, and SciPipe\cite{lampa2019scipipe}. These solutions focus on fields such as Bioinformatics, Big Data, and Machine Learning and address Kubernetes bias to service execution. Because of Kubernetes, they can provide efficient scheduling, automatic recovery, horizontal scalability, and resource monitoring. Most of these solutions do not provide provenance data capture or are limited to versioning and history like Pachyderm or auditing logs like SciPipe. Besides that, their adoption of Kubernetes still makes them less suitable for HPC environments\cite{balis2022auto,zhou2021container}.

To improve the execution of containerized workflows with notebooks, Manne et al.\cite{manne2022chex} address the multi-version replay problem with containers. CHEX is a tool focused on application replay efficiency and aims at containerizing multiple versions of software applications. CHEX records the computational state at a specific program location, named, checkpoints, but it avoids storing a large number of checkpoints through the sharing of common computations. It tackles an NP-Hard problem mainly for notebook-based (REPL) code in machine learning, but it does not address scientific workflows or their deployment in HPC. FLINC\cite{ahmad2020prov} also uses containers to make reproducible experiments that are executed via notebooks. To keep those notebooks interactive, FLINC captures provenance to trace the behavior of the notebook so the container will allow the same behavior. However, adopting a containerized notebook may conflict with the HPC execution environment.

Some scientific workflow management systems (SWfMS)\cite{gerlach2014skyport,filgueira2016asterism,vahi2019custom} have integrated containers into their solutions, providing container and workflow orchestration, significantly reducing the amount of work when compared to manual deployment. However, SWfMSs significantly degrade container-based workflow performance\cite{shan2023kubeadaptor} and do not support multiple deployment strategies for containerization. Integration strategies for containers and workflow systems are proposed by Zheng and Thain\cite{zheng2015integrating} that integrate Docker with Makeflow and Work Queue. While Makeflow is a command-line tool for running data-intensive scientific workflows on various distributed systems, Work Queue is a lightweight execution engine for distributed systems. Thus, the work of Zheng and Thain\cite{zheng2015integrating} supports creating the containerized environment, and their strategies (coarse and fine-grained) vary according to the amount of control the container is given in the workers of Makeflow, but without considering issues related to provenance.

Wofford et al.\cite{wofford2022complete} propose the definition of requirements for capturing the provenance of HPC applications and the issues related to hardware metadata capture. They also propose the design and implementation of a container-based provenance capture system, without discussing workflow containerization. This work highlights the importance and challenges of provenance data capture in container-based applications, including container metadata. One of the challenges is related to the context layers of the application execution, like the context of the HPC environment, the context of containerized application components, and the provenance of the application itself. To tackle this challenge Lim et al.\cite{lim2021secure} propose saBPF (secure audit BPF),  a lightweight system-level auditing framework for container-based cloud environments. It records all container-triggered activity with ProvBPF, a provenance capture mechanism that captures provenance at the thread granularity, recording information such as security context, namespace, and performance metrics. Their provenance is focused on container auditing for security rather than provenance for workflow reproducibility, thus saBPF is complementary to \texttt{ProvDeploy}.

Sciunits\cite{that2017sciunits} is a tool based on provenance for repetition and software reuse. Similar to ReproZip, Sciunits creates, stores, and executes a reusable research object that is a Docker image created from function calls that occurred during the original execution. This research object includes the data consumed and produced by the software application, documentation, and provenance data, so it can be used to repeat the execution in different environments and tools, such as CHEX\cite{manne2022chex}. \texttt{ProvDeploy} also generates research objects, however, SciUnits does not support workflow deployment. Another approach that combines containers and provenance is proposed by Nolte and Wieder\cite{nolte2022realising} which presents a data lake architecture based on FAIR digital objects, which are implemented with containers that provide retrospective provenance and are independent of workflow engine and computational environment. This approach is focused on provenance services needing a workflow execution platform for deployment. It can be considered complementary to \texttt{ProvDeploy}'s architecture as the role of its provenance capture provider. The work of Modi et al.\cite{modi2023querying} also proposes an approach for provenance capture with containers. They developed a tool called Kondo, that, using the provenance previously captured, can determine the files and data that are not accessed and then debloat the containers. While Kondo does not directly support workflow deployment, it represents a significant advance in the data management challenges of containerized workflows and can serve as a complement to \texttt{ProvDeploy}.

Satapaphy et al.\cite{satapathy2023disprotrack} discuss the lack of provenance data capture for microservice applications, and propose DisProTrack\cite{satapathy2023disprotrack} for capturing provenance from microservices in an integrated way, handling parallel calls inherent to microservices. DisProTrack does not have to be installed on other containers and can work from an extra bundle that can handle other container calls. Even though DisProTrack does not support containerized workflow deployment, it presents interesting ideas for container provenance capture.

PROV-CRT\cite{ahmad2020prov} is an approach that aims at tracking provenance computations from the container image during its build processes for auditing and repetition. This approach simplifies container management tasks and container content classification, but, it does not provide ways to capture data during the execution of the generated image and the granularity of the data is system call level which makes its provenance data hard to analyze for reproducibility.

PROV-IO$^+$\cite{han2024prov} stands out as a W3C PROV-compliant provenance tracking framework designed to support both containerized and non-containerized workflow execution across various platforms, including HPC (e.g., clusters and supercomputers) and the cloud. This tool focuses on addressing diverse provenance requirements for scientific data within HPC systems, particularly those traced through I/O operations. For containerized execution, PROV-IO$^+$ provides a \textit{containerizer} tool, enabling the creation of a container image encompassing the entire workflow along with PROV-IO$^+$, condensed into a Docker container image. This image is subsequently converted to Singularity format for execution with PROV-IO$^+$. The experiments presented in the paper show that PROV-IO$^+$ impact on performance is insignificant. In their work, they also propose different types of provenance to be collected, unlike \texttt{ProvDeploy}, their provenance model, though extensible, does not record container metadata, which later can hinder the identification and reproduction of the images used, they also do not deploy multiple containerization strategies.

The work of Olaya et al.\cite{olaya2022building} is the closest approach to \texttt{ProvDeploy}. They present ContainerEnv for containerized workflow execution with provenance data capture using Singularity/Apptainer technology. The provenance is captured in each container of the workflow and later integrated by a Jupyter interface to be used in a \textit{post-mortem} provenance data analysis. ContainerEnv eases the reproducibility, traceability, and explainability of containerized workflows. The authors evaluated ContainerEnv with a real scientific workflow, exploring different configurations, and concluded that a fine-grained strategy is the best option for capturing metadata and ensuring traceability. Unlike \texttt{ProvDeploy}, ContainerEnv only supports a fine-grained workflow containerization strategy. The flexibility of the fine-grained strategy comes with the price of managing multiple container instances during the workflow execution. LandLord\cite{shaffer2023landlord} is an algorithm that focuses on reducing container sprawl, by combining user requests into a specification and exploring existing images that are compatible with the specifications instead of creating multiple container images with similar specifications. Our experiments with \texttt{ProvDeploy} show that adopting a hybrid strategy may address distinct workflow execution goals. For example, when performance is the most important execution criterion, a coarse-grained strategy could be a better choice. \texttt{ProvDeploy} suggests container configurations and allows the user to decide which strategy will benefit most from the environment and qualitative criteria, such as portability and component reuse.

\section{\texttt{ProvDeploy}: Assisting the deployment of containerized scientific workflows in HPC}\label{provdeploysec}

\texttt{ProvDeploy}  is a lightweight framework designed to ease the deployment of containerized scientific workflows in HPC environments while incorporating provenance services. Unlike the current mainstream approaches \cite{lampa2019scipipe,novella2019container,bisong2019kubeflow}, \texttt{ProvDeploy} automates the deployment of a container for provenance services that are integrated with the workflow. This allows users to monitor their workflows at runtime, aiding in debugging and enabling parameter changes during execution (if the workflow allows for adaptations). \texttt{ProvDeploy} allows the user to choose between several available provenance services, but only one can be set as default and is used during workflow execution. The chosen provenance service may require deploying containers with a database management system (DBMS), \textit{e.g.}, MonetDB, PostgreSQL, \textit{etc}. \texttt{ProvDeploy} receives the following information as input: (i) the workflow specification (\textit{i.e.}, in a JSON file), (ii) the datasets to be processed, and (iii) a catalog containing information about available container images that can be deployed using a containerization strategy of the user preference. 

The workflow specification contains the description of the workflow, its scripts, the containerization strategy, \textit{i.e.}, the container images involved including the provenance service, and the corresponding workflow activities. \texttt{ProvDeploy}, then, deploys the containers in a specific environment (\textit{e.g.}, a cluster, or the cloud). \texttt{ProvDeploy} can also support the execution of the defined strategy. If the strategy defined is coarse-grained, \texttt{ProvDeploy} will start the single container image. If the strategy is hybrid or fine-grained, \texttt{ProvDeploy} initially launches and tests containers imperative to the provenance stack, subsequently configuring volumes and binding paths as needed. In this manuscript, we define hybrid strategies as clustering certain components of the execution while separating others into different containers, and always resulting in at least two containers. Following the workflow specification, \texttt{ProvDeploy} sequentially activates container images corresponding to each workflow activity, complying with the specified order. As activities conclude within a container, \texttt{ProvDeploy} proceeds to initiate the subsequent container along with its associated activities. If no provenance service is referenced in the workflow specification, \texttt{ProvDeploy} will start the one that is set as default. 

The same workflow may have multiple viable containerization strategies, and \texttt{ProvDeploy} records data from the containerized execution, which we call container provenance that includes the containerization strategy, the execution environment, the images used, and the execution time for posterior analysis of each strategy, allowing users to investigate the impact of each containerization strategy in the same workflow. It also provides information for repeating the execution and reproducing the container images used if the original images become unavailable. Different from other approaches, by adopting a provenance service, \texttt{ProvDeploy} allows users to have the provenance from the workflow captured according to the user's preferences, easing analysis. In this manner, \texttt{ProvDeploy} can provide two levels of provenance: workflow provenance and container provenance, where the provenance service captures workflow provenance and \texttt{ProvDeploy} captures container provenance that complements the workflow provenance without requiring the adoption of a container-aware provenance service. 

We emphasize that \texttt{ProvDeploy} refrains from controlling parallel and distributed execution. In scenarios where the workflow inherently involves parallel activities, they unfold within the container independently, without interference from \texttt{ProvDeploy}. Upon completion, \texttt{ProvDeploy} generates a research object \cite{bechhofer2010research} as output. This research object encapsulates all the data, metadata, libraries, workflow and container provenance, and dependencies used in the execution of the workflow, towards reproducibility. However, it is important to note that this research object does not capture the specific characteristics of the HPC environment in which the workflow is executed. Instead, it focuses on including the necessary software stack to re-execute the workflow, the selected provenance service, and its provenance database. The research object generated by \texttt{ProvDeploy} is a coarse-grained image that represents the execution, and its purpose is to share the workflow and its results for repetition and verification.

\texttt{ProvDeploy} allows for different provenance services and their addition to the \textit{Catalog} is a separate functionality that receives as input a JSON file with a description of the provenance service image including its tag, repository, hash, volumes, ports, start command, software stack and images that it relies on. For instance, in the current version, \texttt{ProvDeploy} sets DfAnalyzer as the default provenance service, which uses MonetDB as the underlying DBMS and FastBit for data indexing to store the captured data. Suppose we have to add DfAnalyzer as a provenance service, MonetDB and FastBit images would also have to be addressed in the JSON file for DfAnalyzer, and be previously present in the \textit{Catalog}.

Container images used by \texttt{ProvDeploy} can be obtained from various public registries such as DockerHub, Binder, NGC (Nvidia GPU Cloud), \textit{etc}. By leveraging these public container images, our goal is to minimize container build time and assist users in exploring alternative containerization strategies. It is worth mentioning that \texttt{ProvDeploy} is not designed to replace automatic container deployment tools like Kubernetes. Rather, it can be used alongside Kubernetes since \texttt{ProvDeploy} determines which containers will be included in a Kubernetes cluster. The source code for \texttt{ProvDeploy} is available on Bitbucket, accessible through the following URL: \url{https://bitbucket.org/lilianeKunstmann/provdeploy/}. \texttt{ProvDeploy} is an open-source project, which means that its source code is freely available for viewing, modifying, and distributing.

\begin{figure}
      \centering
      \includegraphics[width=.93\textwidth]{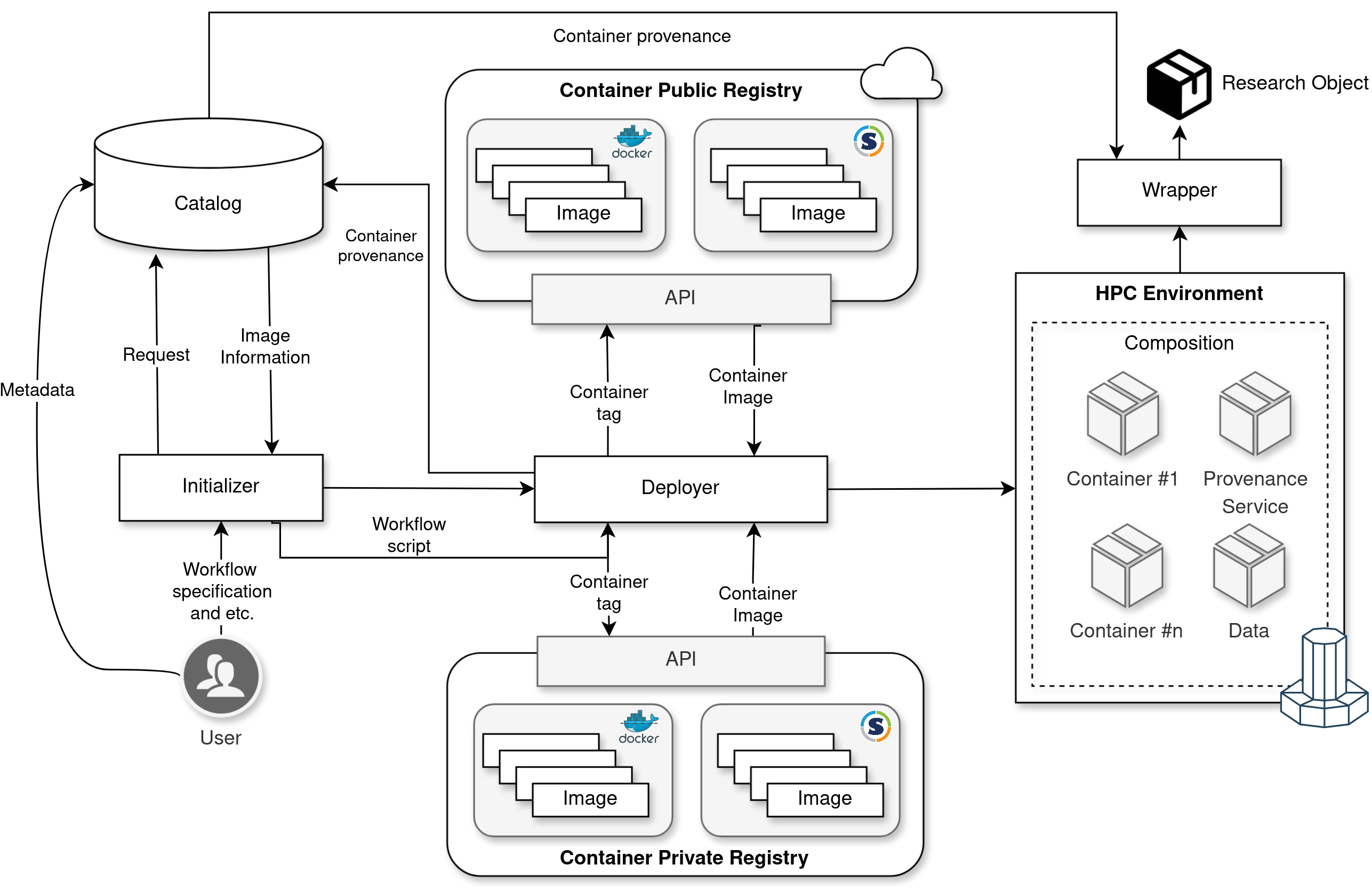} %
      \caption{Architecture of \texttt{ProvDeploy}.}
      \label{fig:arq} 
\end{figure}

The architecture of \texttt{ProvDeploy} is depicted in Figure \ref{fig:arq}, and it consists of four main components: (i) Catalog, (ii) Initializer, (iii) Deployer, and (iv) Wrapper. The \textit{Catalog} serves as a database that stores metadata related to container images that can be deployed by \texttt{ProvDeploy}. It includes information such as the image tag, registry, description, definition file (\textit{e.g.}, dockerfiles, recipes), deployment instructions, and requirements (\textit{e.g.}, volume creation, public ports). The \textit{Catalog} also automatically stores container provenance from the execution of different strategies. This provenance is focused on allowing the user to easily identify the images used, port, drivers, volumes, and bind paths that were set, the provenance service used, and the containerization strategy applied to a workflow. The images present in the \textit{Catalog} are added by the user with local, private, or public images that are going to be pulled when the \textit{Deployer} starts execution. Additionally, the \textit{Catalog} holds details about available provenance services compatible with \texttt{ProvDeploy} (\textit{e.g.}, DfAnalyzer \cite{DBLP:journals/softx/SilvaCGCOCVM20}, noWorkflow \cite{DBLP:journals/pvldb/PimentelMBF17}), as well as DBMS container images (\textit{e.g.}, MonetDB, PostgreSQL, MySQL, \textit{etc}.). This component also manages metadata about versions of the container configuration file (\textit{e.g}, dockerfile, recipe).

The \textit{Initializer} is the user interface with \texttt{ProveDeploy}, it allows the addition of new images to the \textit{Catalog}, including provenance services, sends the workflow to be deployed and allows provenance data, both from the container and the workflow, to be accessed during workflow execution. It queries the \textit{Catalog} to learn if the images required by the workflow are available and their requirements. Some provenance services, such as DfAnalyzer, require instrumentation, which is inserting provenance calls in the workflow script. The \textit{Initializer} assumes that the workflow script is already instrumented to work with the selected provenance service and sends it to the \textit{Deployer}. 

The \textit{Deployer} is the component of \texttt{ProvDeploy} that allows multiple strategies, it follows the strategy described in the workflow specification file and sets up the calls for containers to start and stop according to the workflow activities. The \textit{Deployer} configures the environment where the user's workflow and the chosen provenance service are deployed and can be executed.  The \textit{Deployer} is also responsible for recording the containerization strategy along with images used and their execution time, which helps users find the images used for the execution of some activities. At the end of the execution, the \textit{Deployer} invokes the \textit{Wrapper} to generate a research object related to the workflow execution. The container provenance captured by \texttt{ProvDeploy} is included in the research object by the \textit{Wrapper}, so, even though the research object does not fully represent execution, the container provenance can provide complementary information. This architecture empowers scientists to leverage their resources when executing workflows, seamlessly integrating with the containerization strategy deployed by \texttt{ProvDeploy}. \texttt{ProvDeploy} architecture contributes with its workflow and container provenance data capture and the deployment of hybrid containerization strategies.

\section{\texttt{ProvDeploy} in action}\label{avexp}

This section presents how \texttt{ProvDeploy} manages different containerization strategies with scientific workflows. We use a scientific machine learning workflow named DenseED \cite{freitas2021encoder}. We evaluate the advantages and disadvantages of using different containerization strategies for DenseED in terms of performance using GPUs and CPUs. We have conducted experiments employing three deployment strategies, each configuring DenseED and the provenance components in containers in multiple combinations. Table \ref{tab:descCenario} explains the characteristics and goals of each deployment strategy.

\begin{table}[h]
\centering
\caption{Description of the different containerization strategies used.}
\label{tab:descCenario}
\begin{tabular}{|c|l|l|} 
\hline
\textbf{Strategy} & \textbf{Description} & \textbf{Goal} \\ \hline

%primeir linha
Coarse-grained & 
\begin{tabular}[c]{@{}l@{}}A single container containing all \\
dependencies for provenance data \\ 
capture and running DenseED. \end{tabular} & 
\begin{tabular}[c]{@{}l@{}} 
The most common HPC containerization strategy. \\ 
Eases deployment and a one-step execution. Does\\
not have communication between containers and \\
an instance is enough to do the complete execution. \end{tabular}\\  \hline

%segunda Linha
Partial modular
& \begin{tabular}[c]{@{}l@{}} 
Two containers: A container with  \\
DenseED and a second container   \\
with the DBMS and provenance   \\ 
data, and the provenance service;\end{tabular}& 
\begin{tabular}[c]{@{}l@{}}
Meets the possibility of reusing the DenseED \\
container isolated and getting it from a public \\ 
registry without applying any changes. It isolates \\
data management actions, improving data sharing \\ 
and avoid interfering in the workflow execution.\end{tabular} %software are already containerized allowing \\the reuse of images.  
\\ \hline
%Terceira linha
Provenance modular &
\begin{tabular}[c]{@{}l@{}}
Three containers: the first with  \\  
DenseED, the second with the \\ 
provenance service, and the third \\
with the  DBMS  and provenance \\ data.\end{tabular}                                         
& \begin{tabular}[c]{@{}l@{}}
This case has the same goals as the Partial modular \\ 
and adds the benefits of using images from public \\
registries for deploying provenance and executing \\ 
DenseED. Saves time on image building. It also \\ 
isolates data, so it can be easily shared and swapped, \\
like the provenance service.\end{tabular}  \\ \hline
%\multicolumn{1}{c}{}& \multicolumn{1}{l}{}  & \multicolumn{1}{l}{}                                            
\end{tabular}
\end{table}

\subsection{Case Study - DenseED}

\begin{figure}[h]
    \centering
    \includegraphics[width=0.62\textwidth]{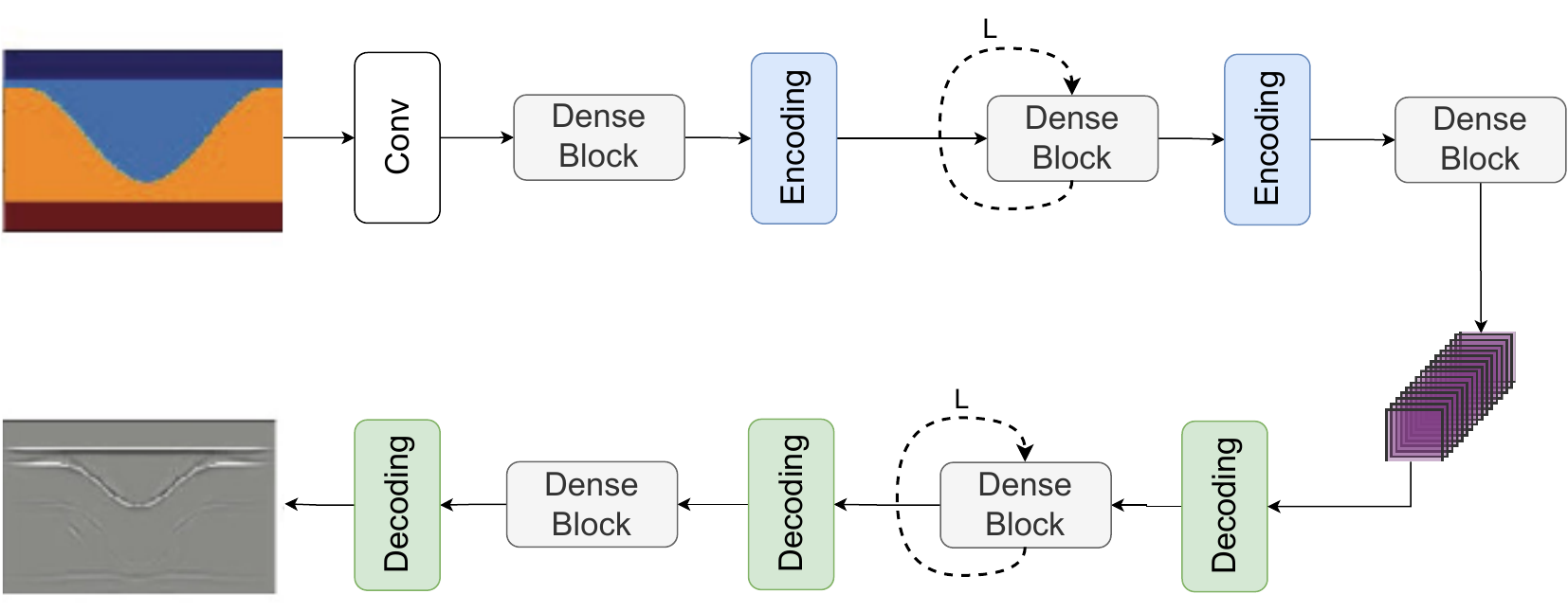}
    \caption{DenseED architecture \cite{freitas2021encoder}.}
    \label{fig:denseed}
\end{figure}

DenseED is a scientific ML workflow introduced by Freitas et al.\cite{freitas2021encoder}. DenseED's architecture is based on a Physics-guided convolutional neural network (CNN) as defined by \cite{zhu2018bayesian}. It comprises convolutional layers and dense blocks, following an encoder-decoder neural network arrangement to handle the potential high-dimensionality of inputs and outputs. DenseED leverages the Physics-guided CNN as a surrogate model for the computationally intensive and time-consuming Reverse Time Migration (RTM) calculations, facilitating quantification of uncertainties \cite{freitas2021encoder}. Traditionally, quantifying the uncertainties in RTM involves solving equations for each probability distribution, which is unfeasible. DenseED aims to replace the calculation of the RTM equations with a Physics-guided trained model. The architecture of DenseED is shown in Figure \ref{fig:denseed}, which takes velocity fields as inputs and generates seismic images with uncertainties as the output. 

DenseED is CPU-intensive, as it processes a substantial input consisting of 200,000 velocity fields. From this pool, it randomly selects 10\% and splits it for training and testing, computes metrics such as standard deviation and mean, saves results into files, and proceeds to build, train, and test the model. Due to the nature of these activities, each activity requires a significant amount of time to start and complete. Our primary goal was to scrutinize and comprehend this specific behavior of DenseED when employed across various containerization strategies with provenance capture on a supercomputer.

It is noteworthy that the version of DenseED utilized in this study is a scaled-down variant, trained with half the minimum number of epochs needed for satisfactory results. This adjustment was intentionally made to expedite observations of the strategy's behavior.

Containers are of great support to DenseED once it relies on Tensorflow. To execute TensorFlow in GPUs we have to match Python, C compiler, Bazel, CUDA, and cuDNN with the available GPU hardware. Matching all these software can become a challenge in HPC facilities. TensorFlow releases are officially available through containers\footnote{\url{https://hub.docker.com/r/tensorflow/tensorflow}}, and registries like NGC provide images optimized for GPUs \footnote{\url{https://catalog.ngc.nvidia.com/orgs/nvidia/containers/tensorflow}}. Using GPUs in Tensorflow containers from NGC will only require enabling GPU usage through flags, but those images are also limited to a range of compatible GPUs, and using them on older GPUs, such as K40, will require downgrading Tensorflow to find a container image that matches the GPU.

\subsection{Environment Setup}

DenseED was deployed with \texttt{ProvDeploy} in the supercomputer Santos Dumont (SDumont) using two different computational nodes a CPU node and a GPU node. SDumont is a cluster with an installed processing capacity of around 5.1 Petaflop/s (5.1 x 1015 float-point operations per second), presenting a hybrid configuration of computational nodes, in terms of the available parallel processing architecture. The CPU node has two CPUs with Intel Xeon E5-2695v2 Ivy Bridge 2.4GHZ processor, 24 cores (12 per CPU), and 64GB DDR3 RAM. The GPU node is part of SDumont expanded partition BullSequana X that has two CPUs with Intel Xeon Skylake  2.1 GHz processor, 48 cores(24 per CPU),  384GB RAM, and four GPUs NVIDIA Volta V100. In both, we used Linux RedHat 7.6 operating system, Singularity 3.8 for the container engine, and for profiling, we used the library sysstat 12.

The experiment procedure consisted of running the DenseED workflow with the DfAnalyzer provenance capture service and its database, MonetDB, in the SDumont. \texttt{ProvDeploy} has been configured to exploit each container deployment strategy described in Table \ref{tab:descCenario}. In this manuscript, we do not explore fine-grained strategy with DenseED because its activities rely on TensorFlow or its dependencies, thus, fine-grained in DenseED would result in container sprawl. Besides, we used an image from NGC that had 6GB, and starting multiple large images would affect performance. We explored partial modular and provenance modular, which are hybrid strategies and seem more coherent with DenseED. There are multiple possibilities for hybrid strategies, but we explored the ones focused on provenance integration.
%\subsection{Quantitative Evaluation}
\subsection{Exploring Different Containerization Strategies}

The evaluation of the three strategies outlined in Table \ref{tab:descCenario} focused on evaluating the overhead of managing several containers by measuring their computational resource usage. We assessed the coarse-grained, partially modular, and provenance modular strategies using metrics such as execution time and CPU consumption. Table \ref{tab:temposcenario} presents the average execution time (\textit{i.e.}, $\overline{x}$) and the standard deviation (\textit{i.e.}, $\sigma$) in minutes across five executions for each strategy, on both CPU and GPU platforms.  

\begin{table}[h]
\centering
\caption{Execution time (in minutes)}
\label{tab:temposcenario}
%\resizebox{\columnwidth}{!}{%
\begin{tabular}{cccrr}
\hline
\multicolumn{1}{|c|}{\textbf{Strategy\textbackslash{}Time(min)}} & \multicolumn{2}{c|}{\textbf{GPU}} & \multicolumn{2}{c|}{\textbf{CPU}} \\ \hline
\multicolumn{1}{|c|}{}                      & \multicolumn{1}{c|}{$\overline{x}$}      & \multicolumn{1}{c|}{$\sigma$} & \multicolumn{1}{c|}{$\overline{x}$}  & \multicolumn{1}{c|}{$\sigma$} \\ \hline
\multicolumn{1}{|c|}{Coarse-grained}        & \multicolumn{1}{c|}{4.214}    & \multicolumn{1}{c|}{0.070}    & \multicolumn{1}{r|}{21.164} & \multicolumn{1}{r|}{0.122} \\ \hline
\multicolumn{1}{|c|}{Partially modular }   & \multicolumn{1}{c|}{4.103}    & \multicolumn{1}{c|}{0.088}    & \multicolumn{1}{r|}{21.514} & \multicolumn{1}{r|}{0.238} \\ \hline
\multicolumn{1}{|c|}{Provenance modular}    & \multicolumn{1}{c|}{4.142}    & \multicolumn{1}{c|}{0.089}    & \multicolumn{1}{r|}{20.711} & \multicolumn{1}{r|}{0.143} \\ \hline
\end{tabular}%
\end{table}

By analyzing Table \ref{tab:temposcenario} we can see a variation that seems negligible in execution times across the different strategies. Consequently, we conducted hypothesis tests for CPU and GPU. In the GPU setting, performing the one-way ANOVA test with $alpha = 5\%$ in all cases, $p-value>alpha$, accepting the null hypothesis, therefore, there are no significant differences between the presented strategies. This result may indicate that, in GPUs, the different strategies do not affect performance, providing more autonomy in choosing the strategy according to the needs of the user. In CPUs, when performing the one-way ANOVA test, there was a significant difference in $\overline{x}$ of the three strategies. Post-hoc analysis with Bonferroni correction, with $alpha = 1,66\%$ in all cases, $p-value<alpha$, rejecting the null hypothesis, therefore, there are significant differences between the presented strategies. Notably, the partially modular strategy emerges as the most time-consuming in CPUs, which can be attributed to bottlenecks in the execution, where one container can be overloaded with distinct tasks, such as executing DenseED and retrieving data or handling provenance calls and persisting and retrieving data.

Even though the coarse-grained strategy also has multiple processes executing within the same container, we expected it to perform best in every aspect since it operates without sharing resources with other containers and avoids handling calls from outside of the container. On the opposite side, we expected provenance modular to perform worst since it starts multiple containers, and requires handling multiple requests. However, in provenance modular, each container is dedicated to a specific task and none of them receives requests for diverse tasks, and this seems to be better for performance \cite{zheng2015integrating}. We aimed to investigate whether these differences could be detected in resource usage.

\begin{figure*}
      \centering
      \includegraphics[width=.95\textwidth]{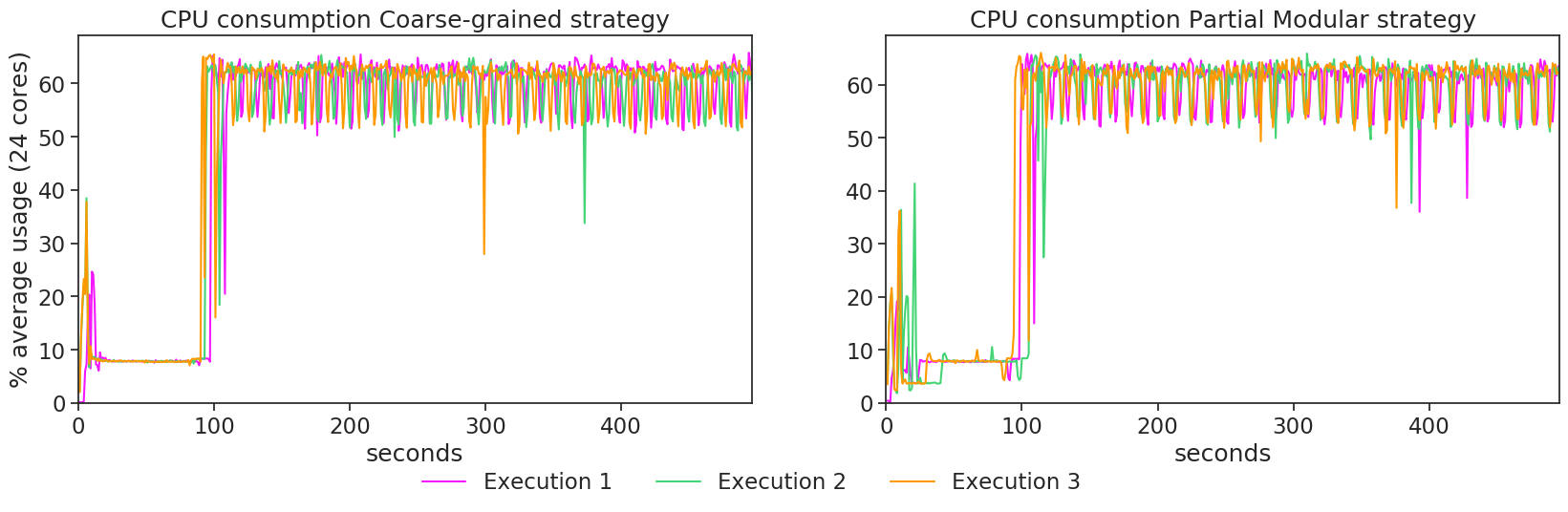} %
       %\vspace{-1em}
      \caption{CPU consumption for coarse-grained and partial modular strategies.}
      \label{fig:comp12sd} 
     
\end{figure*}

\begin{figure*}
\vspace{-1em}
      \centering
      \includegraphics[width=.50\textwidth]{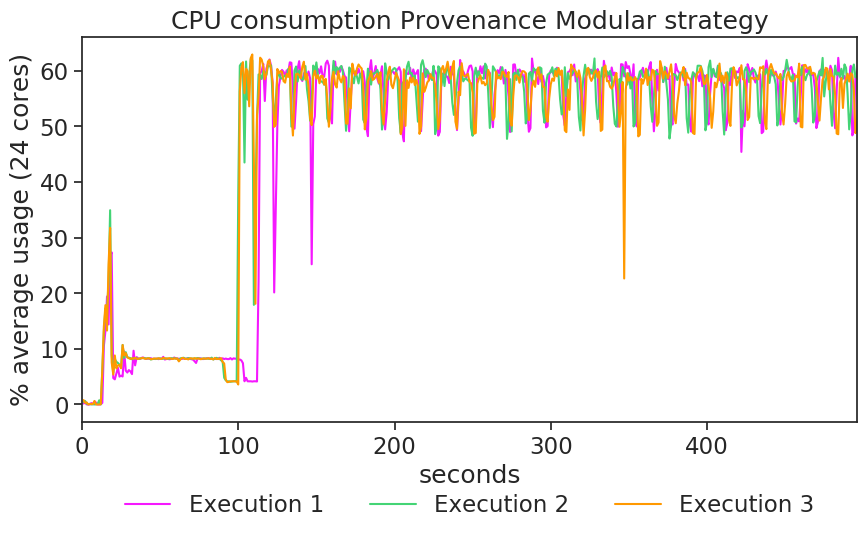} %
      %\vspace{-1em}
      \caption{CPU consumption for provenance modular strategy.}
      \label{fig:comp5sd} 
      
\end{figure*}

We collected the average CPU consumption per second at SDumont. For better visualization, we present the first 500 seconds of three executions for each strategy in Figures \ref{fig:comp12sd} and  \ref{fig:comp5sd}. This average is calculated throughout 24 cores, so the usage can reach 100\% in some cores and hit values close to 0\% on other cores. In the first 100 seconds, we can observe DenseED processing the input dataset by selecting random points for both training and testing. Then it calculates metrics, such as mean and standard deviation, over these data and saves the results in different files. Subsequently, the DenseED workflow proceeds to train and validate the neural network, execute the validation phase, and store predictions along with the Root Mean Square Error (RMSE) for both the training and validation phases. The training, testing, and validation stages are CPU-intensive, resulting in 100\% CPU usage across different cores, while overall CPU consumption rarely drops to 0\%. The average CPU consumption for DenseED is 55\% for the coarse-grained strategy, 56\% for the partial modular strategy, and 52\% for the provenance modular strategy.

The training process starts in approximately 100 seconds. Although provenance modular incurs a slight delay, its CPU consumption is marginally lower than that of coarse-grained and partial modular strategies. This is because it avoids overloading the CPU with multiple activities from the same containers. Additionally, when containers are idle in the CPU, more resources are available for active containers. Considering CPU consumption and execution time the provenance modular strategy emerges as an attractive alternative for DenseED on CPUs. 

The differences detected may become even more significant in larger workflows, so exploring strategies besides coarse and fine-grained can be beneficial to workflow execution. Another observation was the fact that depending on hardware specifications, the performance of the strategies may differ. \texttt{ProvDeploy} can support users in the exploration of hybrid strategies across HPC environments. 

\section{Conclusion}\label{sec:conclusao}

The adoption of software containers as a way to ease workflow deployment is a reality, both in academia and in industry. Despite the container concept representing a breakthrough, there are still challenges in its use in HPC environments and when running scientific workflows that require monitoring and reproduction services, such as capturing provenance data. The user faces defining a containerization strategy to host the main workflow, the consumed and produced data, and the provenance capture service, which is a non-trivial task. A poor decision for containerization strategy impacts the execution time and financial cost or complicates reusing and sharing different data or components. Thus, in this work, we present \texttt{ProvDeploy}, a framework that supports the deployment of containerized workflows. \texttt{ProvDeploy} aims at helping to investigate different containerization strategies to find the most suitable strategy for a given workflow. We evaluated three containerization strategies with DenseED using \texttt{ProvDeploy} on SDumont CPUs and GPUs. The CPU results show that there are significant differences between strategies, the use of multiple containers can impact execution, but it is not an isolated factor, and for DenseED the suggested strategy is provenance modular, a hybrid strategy. In GPUs, as there is no statistical difference between the presented strategies, we could suggest that a hybrid strategy, like also provenance modular or partial modular, would suit DenseED better due to its flexibility and reuse potential over time. Those results evidence that limiting the implementation of a pre-defined containerization strategy can result in a lack of flexibility or execution overhead. This confirms the benefits of using \texttt{ProvDeploy} that allows for multiple options and, using a provenance database to recommend containerization strategies based on previous executions. As future work, we intend to include in \texttt{ProvDeploy} a module for recommending and evaluating the workflow submitted by the user to classify it and recommend the best execution strategy based on previous executions of workflows with similar characteristics.

\section*{Acknowledgments}
This was supported in part by CNPq, FAPERJ, and Coordenação de Aperfeiçoamento de Pessoal de Nível Superior - Brazil (CAPES) - Finance Code 001.
%Bibliography
\bibliographystyle{unsrt}  
\bibliography{references}

\begin{thebibliography}{10}

\bibitem{DBLP:journals/access/EliaFA21}
Donatello Elia, Sandro Fiore, and Giovanni Aloisio.
\newblock Towards {HPC} and big data analytics convergence: Design and experimental evaluation of a {HPDA} framework for escience at scale.
\newblock {\em {IEEE} Access}, 9:73307--73326, 2021.

\bibitem{yuan2020bioinformatics}
David~Yu Yuan and Tony Wildish.
\newblock Bioinformatics application with kubeflow for batch processing in clouds.
\newblock In {\em HPDC}, pages 355--367. Springer, 2020.

\bibitem{freire2008provenance}
Juliana Freire, David Koop, Emanuele Santos, and Cl{\'a}udio~T Silva.
\newblock Provenance for computational tasks: A survey.
\newblock {\em Computing in science \& engineering}, 10(3):11--21, 2008.

\bibitem{DBLP:conf/eScience/OcanaSOM15}
Kary~ACS Oca{\~n}a, Vítor Silva, Daniel {de Oliveira}, and Marta Mattoso.
\newblock Data analytics in bioinformatics: Data science in practice for genomics analysis workflows.
\newblock In {\em IEEE e-Science}, pages 322--331. IEEE, 2015.

\bibitem{DBLP:journals/cluster/GuedesJODO20}
Thaylon Guedes, Leonardo~A Jesus, Kary~ACS Oca{\~n}a, Lucia Drummond, and Daniel {de Oliveira}.
\newblock Provenance-based fault tolerance technique recommendation for cloud-based scientific workflows: a practical approach.
\newblock {\em Cluster Comp.}, 23(1):123--148, 2020.

\bibitem{DBLP:journals/tpds/HarrellMM22}
Stephen~Lien Harrell, Scott Michael, and Carlos Maltzahn.
\newblock Advancing adoption of reproducibility in {HPC:} {A} preface to the special section.
\newblock {\em {IEEE} Trans. Par. Dist. Syst.}, 33(9):2011--2013, 2022.

\bibitem{DBLP:conf/fogiot/StruharBAP20}
V{\'{a}}clav Struh{\'{a}}r, Moris Behnam, Mohammad Ashjaei, and Alessandro~Vittorio Papadopoulos.
\newblock Real-time containers: {A} survey.
\newblock In {\em Fog-IoT}, volume~80 of {\em OASIcs}, pages 7:1--7:9, 2020.

\bibitem{DBLP:journals/cluster/LiuG22}
Peini Liu and Jordi Guitart.
\newblock Performance characterization of containerization for {HPC} workloads on infiniband clusters: an empirical study.
\newblock {\em Clust. Comput.}, 25(2):847--868, 2022.

\bibitem{priedhorsky2021minimizing}
Reid Priedhorsky, R~Shane Canon, Timothy Randles, and Andrew~J Younge.
\newblock Minimizing privilege for building hpc containers.
\newblock In {\em Proceedings of the International Conference for High Performance Computing, Networking, Storage and Analysis}, pages 1--14, November 2021.

\bibitem{canon2020role}
R~Shane Canon.
\newblock The role of containers in reproducibility.
\newblock In {\em 2020 2nd International Workshop on Containers and New Orchestration Paradigms for Isolated Environments in HPC (CANOPIE-HPC)}, pages 19--25. IEEE, December 2020.

\bibitem{wofford2022complete}
Quincy Wofford, James Hurd, Hugh Greenberg, Patrick~G Bridges, and James Ahrens.
\newblock Complete provenance for application experiments with containers and hardware interface metadata.
\newblock In {\em 2022 IEEE/ACM 4th International Workshop on Containers and New Orchestration Paradigms for Isolated Environments in HPC (CANOPIE-HPC)}, pages 12--24. IEEE, 2022.

\bibitem{schulz2016use}
Wade~L Schulz, Thomas~JS Durant, Alexa~J Siddon, and Richard Torres.
\newblock Use of application containers and workflows for genomic data analysis.
\newblock {\em Journal of pathology informatics}, 7(1):53, 2016.

\bibitem{elisseev2022multiscale}
Vadim Elisseev, Robert Manson-Sawko, Carlos Pe{\~n}a-Monferrer, Guido Lupieri, Michael Seaton, Gianluca Boccardo, Jan-Willem Handgraaf, Ilian Todorov, Daniele Marchisio, and Adam Kowalski.
\newblock Multiscale scientific workflows on high-performance hybrid cloud.
\newblock In {\em 2022 IEEE/ACM 4th International Workshop on Containers and New Orchestration Paradigms for Isolated Environments in HPC (CANOPIE-HPC)}, pages 1--11. IEEE, 2022.

\bibitem{ahmad2022reproducible}
Raza Ahmad, Naga~Nithin Manne, and Tanu Malik.
\newblock Reproducible notebook containers using application virtualization.
\newblock In {\em 2022 IEEE 18th International Conference on e-Science (e-Science)}, pages 1--10. IEEE, 2022.

\bibitem{kleinsteuber2024managing}
Erik Kleinsteuber, Tarek Al~Mustafa, Franziska Zander, Birgitta K{\"o}nig-Ries, and Samira Babalou.
\newblock Managing provenance data in knowledge graph management platforms.
\newblock pages 1--10. Springer, 2024.

\bibitem{abbas2022paced}
Mashal Abbas, Shahpar Khan, Abdul Monum, Fareed Zaffar, Rashid Tahir, David Eyers, Hassaan Irshad, Ashish Gehani, Vinod Yegneswaran, and Thomas Pasquier.
\newblock Paced: Provenance-based automated container escape detection.
\newblock In {\em 2022 IEEE International Conference on Cloud Engineering (IC2E)}, pages 261--272. IEEE, 2022.

\bibitem{chen2021clarion}
Xutong Chen, Hassaan Irshad, Yan Chen, Ashish Gehani, et~al.
\newblock Clarion: Sound and clear provenance tracking for microservice deployments.
\newblock In {\em USENIX Security}, pages 3989--4006, 2021.

\bibitem{gerhardt2017shifter}
Lisa Gerhardt, Wahid Bhimji, Shane Canon, Markus Fasel, Doug Jacobsen, Mustafa Mustafa, Jeff Porter, and Vakho Tsulaia.
\newblock Shifter: Containers for hpc.
\newblock In {\em Journal of physics: Conference series}, volume 898, page 082021. IOP Publishing, 2017.

\bibitem{priedhorsky2017charliecloud}
Reid Priedhorsky and Tim Randles.
\newblock Charliecloud: unprivileged containers for user-defined software stacks in hpc.
\newblock In {\em Proceedings of the International Conference for High Performance Computing, Networking, Storage and Analysis}, SC '17, New York, NY, USA, 2017. Association for Computing Machinery.

\bibitem{kurtzer2017singularity}
Gregory~M Kurtzer, Vanessa Sochat, and Michael~W Bauer.
\newblock Singularity: Scientific containers for mobility of compute.
\newblock {\em PloS one}, 12(5):e0177459, 2017.

\bibitem{cito2017empirical}
J{\"u}rgen Cito, Gerald Schermann, John~Erik Wittern, Philipp Leitner, Sali Zumberi, and Harald~C Gall.
\newblock An empirical analysis of the docker container ecosystem on github.
\newblock In {\em 2017 IEEE/ACM 14th International Conference on Mining Software Repositories (MSR)}, pages 323--333. IEEE, 2017.

\bibitem{moreau2013provenance}
Luc Moreau and Paul Groth.
\newblock Provenance: an introduction to prov.
\newblock {\em Synthesis lectures on the semantic web: theory and technology}, 3(4):1--129. {Morgan \& Claypool Publishers}, 2013.

\bibitem{MATTOSO2015100}
Marta Mattoso, Jonas Dias, Kary~ACS Ocana, Eduardo Ogasawara, Flavio Costa, Felipe Horta, Vitor Silva, and Daniel De~Oliveira.
\newblock Dynamic steering of hpc scientific workflows: A survey.
\newblock {\em Future Generation Computer Systems}, 46:100--113, 2015.

\bibitem{DBLP:journals/softx/SilvaCGCOCVM20}
V{\'\i}tor Silva, Vin{\'\i}cius Campos, Thaylon Guedes, Jos{\'e} Camata, Daniel de~Oliveira, Alvaro~LGA Coutinho, Patrick Valduriez, and Marta Mattoso.
\newblock Dfanalyzer: Runtime dataflow analysis tool for computational science and engineering applications.
\newblock {\em SoftwareX}, 12:100592, 2020.

\bibitem{balis2022auto}
Bartosz Bali{\'s}, Andrzej Bro{\'n}ski, and Mateusz Szarek.
\newblock Auto-scaling of scientific workflows in kubernetes.
\newblock In {\em ICCS}, pages 33--40. Springer, 2022.

\bibitem{zhou2021container}
Naweiluo Zhou, Yiannis Georgiou, Marcin Pospieszny, Li~Zhong, Huan Zhou, Christoph Niethammer, Branislav Pejak, Oskar Marko, and Dennis Hoppe.
\newblock Container orchestration on hpc systems through kubernetes.
\newblock {\em Journal of Cloud Computing}, 10(1):1--14, 2021.

\bibitem{straesserempirical}
Martin Straesser, Andr{\'e} Bauer, Robert Leppich, Nikolas Herbst, Kyle Chard, Ian Foster, and Samuel Kounev.
\newblock An empirical study of container image configurations and their impact on start times.
\newblock {\em IEEE Xplore}, 2023.

\bibitem{DBLP:conf/sigmod/ChirigatiRSF16}
Fernando Chirigati, R{\'{e}}mi Rampin, Dennis~E. Shasha, and Juliana Freire.
\newblock Reprozip: Computational reproducibility with ease.
\newblock In {\em SIGMOD}, pages 2085--2088. ACM, {ACM}, 2016.

\bibitem{that2017sciunits}
Dai~Hai Ton~That, Gabriel Fils, Zhihao Yuan, and Tanu Malik.
\newblock Sciunits: Reusable research objects.
\newblock In {\em 2017 IEEE 13th International Conference on e-Science (e-Science)}, pages 374--383, 2017.

\bibitem{bechhofer2010research}
Sean Bechhofer, David De~Roure, Matthew Gamble, Carole Goble, and Iain Buchan.
\newblock Research objects: Towards exchange and reuse of digital knowledge.
\newblock {\em Nature Proc.}, pages 1--6, 2010.

\bibitem{rampin2018reproserver}
Remi Rampin, Fernando Chirigati, Vicky Steeves, and Juliana Freire.
\newblock Reproserver: making reproducibility easier and less intensive.
\newblock {\em arXiv preprint arXiv:1808.01406}, 2018.

\bibitem{shan2023kubeadaptor}
Chenggang Shan, Yuanqing Xia, Yufeng Zhan, and Jinhui Zhang.
\newblock Kubeadaptor: A docking framework for workflow containerization on kubernetes.
\newblock {\em Future Generation Computer Systems}, 148:584--599, 2023.

\bibitem{novella2019container}
Jon~Ander Novella, Payam Emami~Khoonsari, Stephanie Herman, Daniel Whitenack, Marco Capuccini, Joachim Burman, Kim Kultima, and Ola Spjuth.
\newblock Container-based bioinformatics with pachyderm.
\newblock {\em Bioinformatics}, 35(5):839--846, 2019.

\bibitem{bisong2019kubeflow}
Ekaba Bisong.
\newblock Kubeflow and kubeflow pipelines.
\newblock {\em Building Machine Learning and Deep Learning Models on Google Cloud Platform: A Comprehensive Guide for Beginners}, pages 671--685, 2019.

\bibitem{lampa2019scipipe}
Samuel Lampa, Martin Dahl{\"o}, Jonathan Alvarsson, and Ola Spjuth.
\newblock Scipipe: A workflow library for agile development of complex and dynamic bioinformatics pipelines.
\newblock {\em GigaScience}, 8(5):giz044, 2019.

\bibitem{manne2022chex}
Naga~Nithin Manne, Shilvi Satpati, Tanu Malik, Amitabha Bagchi, Ashish Gehani, and Amitabh Chaudhary.
\newblock Chex: multiversion replay with ordered checkpoints.
\newblock {\em Proceedings of the VLDB Endowment}, 15(6):1297--1310, 2022.

\bibitem{ahmad2020prov}
Raza Ahmad, Yuta Nakamura, Naga~Nithin Manne, and Tanu Malik.
\newblock Prov-crt: Provenance support for container runtimes.
\newblock In {\em TaPP 2020}, pages 1--3, 2020.

\bibitem{gerlach2014skyport}
Wolfgang Gerlach, Wei Tang, Kevin Keegan, Travis Harrison, Andreas Wilke, Jared Bischof, Mark D'Souza, Scott Devoid, Daniel Murphy-Olson, Narayan Desai, et~al.
\newblock Skyport-container-based execution environment management for multi-cloud scientific workflows.
\newblock In {\em 2014 5th International Workshop on Data-Intensive Computing in the Clouds}, pages 25--32. IEEE, 2014.

\bibitem{filgueira2016asterism}
Rosa Filgueira, Rafael~Ferreira Da~Silva, Amrey Krause, Ewa Deelman, and Malcolm Atkinson.
\newblock Asterism: Pegasus and dispel4py hybrid workflows for data-intensive science.
\newblock In {\em 2016 Seventh International Workshop on Data-Intensive Computing in the Clouds (DataCloud)}, pages 1--8. IEEE, 2016.

\bibitem{vahi2019custom}
Karan Vahi, Mats Rynge, George Papadimitriou, Duncan~A Brown, Rajiv Mayani, Rafael~Ferreira da~Silva, Ewa Deelman, Anirban Mandal, Eric Lyons, and Michael Zink.
\newblock Custom execution environments with containers in pegasus-enabled scientific workflows.
\newblock In {\em 2019 15th International Conference on eScience (eScience)}, pages 281--290. IEEE, 2019.

\bibitem{zheng2015integrating}
Charles Zheng and Douglas Thain.
\newblock Integrating containers into workflows: A case study using makeflow, work queue, and docker.
\newblock In {\em WVTDC}, pages 31--38, 2015.

\bibitem{lim2021secure}
Soo~Yee Lim, Bogdan Stelea, Xueyuan Han, and Thomas Pasquier.
\newblock Secure namespaced kernel audit for containers.
\newblock In {\em Proceedings of the ACM Symposium on Cloud Computing}, pages 518--532, 2021.

\bibitem{nolte2022realising}
Hendrik Nolte and Philipp Wieder.
\newblock Realising data-centric scientific workflows with provenance-capturing on data lakes.
\newblock {\em Data Intelligence}, 4(2):426--438, 2022.

\bibitem{modi2023querying}
Aniket Modi, Moaz Reyad, Tanu Malik, and Ashish Gehani.
\newblock Querying container provenance.
\newblock In {\em Companion Proceedings of the ACM Web Conference 2023}, WWW '23 Companion, page 1564–1567, New York, NY, USA, 2023. Association for Computing Machinery.

\bibitem{satapathy2023disprotrack}
Utkalika Satapathy, Rishabh Thakur, Subhrendu Chattopadhyay, and Sandip Chakraborty.
\newblock Disprotrack: Distributed provenance tracking over serverless applications.
\newblock In {\em IEEE INFOCOM 2023-IEEE Conference on Computer Communications}, pages 1--10. IEEE, 2023.

\bibitem{han2024prov}
Runzhou Han, Mai Zheng, Suren Byna, Houjun Tang, Bin Dong, Dong Dai, Yong Chen, Dongkyun Kim, Joseph Hassoun, and David Thorsley.
\newblock Prov-io$^+$: A cross-platform provenance framework for scientific data on hpc systems.
\newblock {\em IEEE Transactions on Parallel and Distributed Systems}, 2024.

\bibitem{olaya2022building}
Paula Olaya, Dominic Kennedy, Ricardo Llamas, Leobardo Valera, Rodrigo Vargas, Jay Lofstead, and Michela Taufer.
\newblock Building trust in earth science findings through data traceability and results explainability.
\newblock {\em IEEE Transactions on Parallel and Distributed Systems}, 34(2):704--717, 2022.

\bibitem{shaffer2023landlord}
Tim Shaffer, Thanh~Son Phung, Kyle Chard, and Douglas Thain.
\newblock Landlord: Coordinating dynamic software environments to reduce container sprawl.
\newblock {\em IEEE Transactions on Parallel and Distributed Systems}, 2023.

\bibitem{DBLP:journals/pvldb/PimentelMBF17}
Jo{\~{a}}o~Felipe Pimentel, Leonardo Murta, Vanessa Braganholo, and Juliana Freire.
\newblock noworkflow: a tool for collecting, analyzing, and managing provenance from python scripts.
\newblock {\em Proc. {VLDB} Endow.}, 10(12):1841--1844, 2017.

\bibitem{freitas2021encoder}
Rodolfo~SM Freitas, Carlos~HS Barbosa, Gabriel~M Guerra, Alvaro~LGA Coutinho, and Fernando~A Rochinha.
\newblock An encoder-decoder deep surrogate for reverse time migration in seismic imaging under uncertainty.
\newblock {\em Computational Geosciences}, 25:1229--1250, 2021.

\bibitem{zhu2018bayesian}
Yinhao Zhu and Nicholas Zabaras.
\newblock Bayesian deep convolutional encoder--decoder networks for surrogate modeling and uncertainty quantification.
\newblock {\em Journal of Computational Physics}, 366:415--447, 2018.

\end{thebibliography}

\end{document}